\documentclass[journal=acs,manuscript=article]{achemso}
\setkeys{acs}{maxauthors=0}
\usepackage{adjustbox}
\usepackage[table]{xcolor} 
\usepackage{url}
\usepackage{amsmath}
\usepackage{amssymb}
\usepackage{bm}  
\usepackage{booktabs}
\usepackage{color}
\usepackage{dcolumn} 
\usepackage{etoolbox}
\usepackage[T1]{fontenc}
\usepackage{framed}
\usepackage{geometry}
\usepackage{graphicx}  
\usepackage{hhline}
\usepackage{hyperref}
\usepackage[utf8]{inputenc}
\usepackage{lipsum}
\usepackage{longtable}
\usepackage[version=3]{mhchem} 
\usepackage{multirow}
\usepackage{rotating}
\usepackage{subfigure,amsmath,verbatim,moreverb}
\usepackage{tabularx}
\usepackage{threeparttable} 
\usepackage{xcolor}

\definecolor{bluegreen}{rgb}{0,0.2,0.8}

\UseRawInputEncoding

\AtBeginEnvironment{align}{\setcounter{subeqn}{0}}
\newcounter{subeqn} %

\newcommand{\Tab}[1]{Table \ref{#1}}

\usepackage{color}

\author{Aditi Singh}
\email{aditisingh4812@gmail.com}
\affiliation{Institute of Physics, Faculty of Physics, Astronomy and Informatics, Nicolaus Copernicus University in Toru\'n,
ul. Grudzi\k{a}dzka 5, 87-100 Toru\'n, Poland}
\author{Ram Dhari Pandey}
\email{pandey123iitian@gmail.com}
\affiliation{Institute of Physics, Faculty of Physics, Astronomy and Informatics, Nicolaus Copernicus University in Toru\'n,
ul. Grudzi\k{a}dzka 5, 87-100 Toru\'n, Poland}
\author{Subrata Jana}
 \affiliation{Institute of Physics, Faculty of Physics, Astronomy and Informatics, Nicolaus Copernicus University in Toru\'n,
ul. Grudzi\k{a}dzka 5, 87-100 Toru\'n, Poland}
\author{Prasanjit Samal}
\affiliation{School of Physical Sciences, National Institute of Science Education and Research, An OCC of Homi Bhabha National Institute, Bhubaneswar 752050, India}
\author{Pawe\l~Tecmer}
\affiliation{Institute of Physics, Faculty of Physics, Astronomy and Informatics, Nicolaus Copernicus University in Toru\'n,
ul. Grudzi\k{a}dzka 5, 87-100 Toru\'n, Poland}
\author{Szymon \'Smiga}
\email{szsmiga@fizyka.umk.pl}
\affiliation{Institute of Physics, Faculty of Physics, Astronomy and Informatics, Nicolaus Copernicus University in Toru\'n,
ul. Grudzi\k{a}dzka 5, 87-100 Toru\'n, Poland}

\title{Frontier Orbital Engineering
in Heteroatom-Doped Prototypical Organic Dyes for Dye-Sensitized Solar
Cells
}

\begin{document}

\begin{abstract}

The computational design of heteroatom-doped organic dyes for dye-sensitized solar cells (DSSCs) remains challenging, as predictive methods must accurately describe long-range charge-transfer (CT) excitations while remaining computationally efficient for systematic materials screening. 
In this work, we investigate the electronic structure and excited-state properties using the range-separated hybrid functional LC-$\omega$PBE in conjunction with linear-response time-dependent density functional theory (TDDFT) within the Tamm-Dancoff approximation (TDA).
We employ a simplified, physically motivated, effective tuning protocol ($\omega_{eff}$) to enable the rapid and reliable screening of electronic properties of organic dyes. Charge-transfer excitation energies and frontier orbital alignment the key factors governing light absorption and electron injection in DSSCs are analyzed through targeted heteroatom (N, O, and B) incorporation into donor-$\pi$-acceptor (D-$\pi$-A) organic dyes. A library of 27 mono-, di-, and tri-doped prototypical organic dyes is designed based on a carbazole donor and a cyanoacrylic acid acceptor through targeted doping at three positions of the $\pi$-bridge or linker. Distinct design trends emerge: electron-rich nitrogen and oxygen dopants increase the HOMO-LUMO gap and blue-shift CT excitations, with nitrogen exhibiting the strongest effect, whereas electron-deficient boron substitution narrows the gap and induces pronounced red shifts. Notably, the BBN-doped dye exhibits the smallest gap and lowest excitation energy, highlighting boron-rich motifs as promising candidates for enhanced solar light harvesting. Overall, this study establishes transferable heteroatom-doping guidelines and introduces an efficient, reliable, and cost-effective tuned DFT-TDDFT framework for high-throughput computational discovery and optimization of DSSC sensitizers.
\end{abstract}

\maketitle

\section{I. Introduction}

\begin{figure*}
    \centering
    \includegraphics[width=0.9\linewidth]{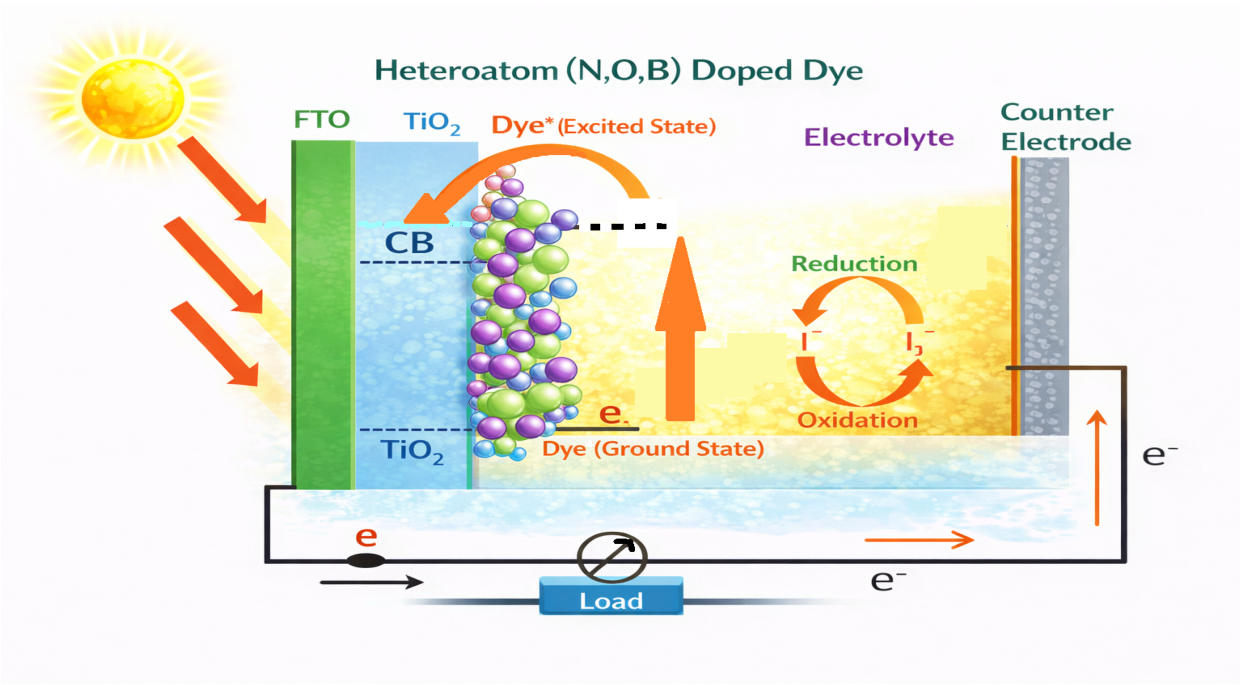}
    \caption{Schematic illustration of the operating principle of DSSCs.}
    \label{fig:solar}
\end{figure*}

Dye-sensitized solar cells (DSSCs) are a distinct class of photovoltaic devices that generate electricity through molecular light harvesting rather than bulk semiconductor absorption. By decoupling light absorption from charge transfer (CT), DSSCs employ organic or organometallic sensitizers anchored to a wide-bandgap semiconductor, most commonly nanocrystalline \ce{TiO2}. This architecture enables low-cost fabrication\cite{GRATZEL2003145}, mechanical flexibility\cite{KREBS2009394}, color tunability\cite{Nazeeruddin2006}, and efficient operation under diffuse or indoor illumination\cite{Mathew2014}, making DSSCs attractive for next-generation photovoltaic applications. \cite{ORegan1991,Gratzel2000,Hagfeldt2010} The performance of a DSSC is governed by a sequence of interfacial CT processes, including photoexcitation of the dye, electron injection into the semiconductor conduction band, dye regeneration by the redox electrolyte, and suppression of charge recombination. Among these steps, efficient and directional charge transfer from the photoexcited dye to the semiconductor is particularly critical. Consequently, achieving high device efficiency requires precise control over CT energetics, excited-state alignment, donor-acceptor separation, and electronic coupling at the dye-semiconductor interface. Even minor molecular modifications have been shown to markedly influence charge-transfer character, electron injection driving forces, and recombination dynamics in DSSCs. \cite{DeAngelis2014} A key molecular design principle for DSSC sensitizers therefore lies in the careful engineering of frontier molecular orbitals, specifically the highest occupied molecular orbital (HOMO) and lowest unoccupied molecular orbital (LUMO). These energetic requirements directly determine the efficiency of charge separation and overall device performance. \cite{Gratzel2005,Freitag2017}\\
Traditional approaches to designing artificial sensitizers rely on costly metal-based complexes like those using ruthenium~\cite{Ru-complexes}, which require complex purification. In contrast, metal-free organic dyes offer many advantages, including a large absorption coefficient,~\cite{energy-homo-lumo, Zhang2016} simple and low-cost synthesis process,~\cite{synthesis2,synthesis,synthesis1} an environmentally friendly approach, and a broad photon spectrum that facilitates efficient charge separation across the dye molecule~\cite{broad-spectrum}. A wide range of metal-free organic dyes have been studied with different configurations such as D-A-$\pi$-A~\cite{D-A-PI-A, Double-A-A}, D-D-$\pi$-A~\cite{D-D-PI-A}, D-$\pi$-A~\cite{D-PI-A}, D-$\pi$-A-A~\cite{D-PI-A-A}, where D and A are donor and acceptor, respectively. Among them, the widely accepted and highly promising approach for molecular design of efficient metal-free organic sensitizers is the donor-$\pi$-bridge-acceptor (D-$\pi$-A) model~\cite{efficient1-D-pi-A,efficient2-D-pi-A,efficient3-D-pi-A}, which allows intramolecular CT. In this framework, the possibility of optimal donor units such as triphenylamine,~\cite{tpp1,tpp2} coumarin,~\cite{coumarin1,coumarin2,coumarin3} carbazole,~\cite{carbazole1,carbazole2} and phenothiazine being the best option due to their strong electron-donating capacity. The $\pi$-conjugated bridge, often composed of different moieties such as perylene,~\cite{perylene1,perylene2} enediyne,~\cite{enediyne} thiophene and fused benzene rings, serves as a crucial linker between the donor and acceptor. Finally, cyanoacrylic acid is the predominant acceptor and anchoring group,~\cite{anchoring-group2,anchoring-group3,anchoring-group4} as it effectively withdraws electrons from the donor via its cyano group (-CN) while its carboxylic acid group (-COOH) binds to the \ce{TiO2} surface. When photons are absorbed by an organic dye, electrons are excited from the HOMO to the LUMO.~\cite{homo-lumo-excitation} To achieve efficient device operation, two key energy level conditions (see Figure~\ref{fig:solar}) must be met.~\cite{energy-homo-lumo} First, the HOMO level of the dye must be significantly below the redox potential (-4.80 eV $I^{-}$/$I^{-}_3$) of the electrolyte for efficient regeneration of the dye. Second, the LUMO energy level of the organic dye should be above the energy level of the conduction band (-4.0 eV \ce{TiO2}) of the semiconductor. This alignment facilitates the injection of photoexcited electrons from the dye's LUMO into the conduction band.\\
Based on our review of the current literature, and to the best of our knowledge, there remains a substantial gap in understanding the electronic structures of carbazole-based systems featuring five-membered, heteroatom-doped $\pi$-bridges.~\cite{unexplored,lena-jctc}

Only a few systems have been studied, leaving this area largely unexplored so far. The limited exploration is primarily caused by several challenges contributing to this difficulty i.e. (i) the complexity and computational cost of the systems, especially with wave function theory (WFT) methods\cite{CTfails, szabo1996modern, kozma2020benchmark, jacquemin2008tddft}; (ii) the constraints of using small basis sets\cite{jacquemin2009extensive, CTfails, thiel,ram-jpca,SINGH2023297} which in context of WFT methods can provide not so reliable predicitons\cite{smallBS}; (iii) the lack of cost-effective approaches that can efficiently and accurately explore the broad chemical space of possible complexes while maintaining predictive accuracy\cite{thiel, unexplored, Curtarolo2013, ram2025frozen, ram-jpca, CTfails, Onida2002electronic, Aditi_orbital}. As a result, the lack of organized datasets and systematic studies have prevented a clear understanding of how doping affects DSSC properties. 

To efficiently sweep a broad chemical space of possible DSSC complexes, our computational workflow combines ground-state Kohn-Sham density functional theory (KS-DFT) 
and linear-response time-dependent DFT (TDDFT) for excited-state properties treatment. The former calculations have been performed using the range-separated hybrid (RSH) functional LC-$\omega$PBE \cite{lcwpbe}, which is well suited for describing CT character due to its correct long-range treatment of exchange energy\cite{Stein2009Reliable, singh2025simplifiedphysicallymotivateduniversally, Mandal2025Simplified} and potential\cite{Kumar24}. In turn, the range-separation (RS) parameter $\omega$ was set using the recently proposed efficient $\omega_{eff}$ tuning protocol \cite{singh2025simplifiedphysicallymotivateduniversally}, which has been shown to provide near-IP-tuning accuracy at a substantially reduced computational cost. This methodology provides a physically motivated, system-dependent starting point that has been empirically validated for CT state prediction\cite{singh2025simplifiedphysicallymotivateduniversally}. Moreover, recent studies have demonstrated that $\omega_{eff}$ serves as an excellent starting point for single-shot Green's function approximation $G_0W_0$ calculations within many-body perturbation theory\cite{new_aditi}. This dual validation within both TDDFT and $G_0W_0$ frameworks substantiates the robustness of $\omega_{eff}$ as a reliable parameter for both predicting properties of existing CT systems and guiding the design of novel donor-acceptor architectures.

By leveraging these methods to address the existing gap in the literature and enable rational materials design, we systematically investigate mono-, di-, and tri-doped organic dyes consisting of a common carbazole donor and a cyanoacrylic acid acceptor.

The paper is organized as follows. Section II provides the methodological background, and Section III presents and discusses the results. Finally, Section IV provides concluding remarks and future perspectives.

\section{II. Methodology}
\label{sec_method}
\subsection{A. Methods}

Ground-state electronic structure calculations were performed using the range-separated hybrid (RSH) density functional LC-$\omega$PBE. \cite{lcwpbe} In this functional, the electron--electron Coulomb interaction is partitioned into short-range (SR) and long-range (LR) components using an error-function formalism,
\begin{equation}
\frac{1}{r} =
\frac{\alpha + \beta\,\mathrm{erf}(\omega r)}{r}
+ \frac{1 - \left[\alpha + \beta\,\mathrm{erf}(\omega r)\right]}{r},
\label{eq:range_separation}
\end{equation}
where $\omega$ in Eq.~\ref{eq:range_separation} denotes the range-separation parameter that governs the onset of the long-range exchange interaction. For LC-$\omega$PBE, the parameters are fixed to $\alpha = 0$ and $\beta = 1$, such that 100\% Hartree--Fock (HF) exchange is recovered in the long-range limit. The corresponding exchange-correlation energy is given by
\begin{equation}
E_{\mathrm{xc}}(\omega) =
E_{\mathrm{x}}^{\mathrm{LR,HF}}(\omega)
+ E_{\mathrm{x}}^{\mathrm{SR,PBE}}(\omega)
+ E_{\mathrm{c}}^{\mathrm{PBE}},
\label{eq:lcwpbe_xc}
\end{equation}
where $E_{\mathrm{x}}^{\mathrm{LR,HF}} (\omega)$ in Eq.~\ref{eq:lcwpbe_xc} represents the long-range HF exchange contribution, $E_{\mathrm{x}}^{\mathrm{SR,PBE}}(\omega)$ corresponds to the short-range PBE exchange term and $E_{\mathrm{c}}^{\mathrm{PBE}}$ denotes the full-range PBE correlation functional. By enforcing exact HF exchange at long range, LC-$\omega$PBE yields an exchange potential with the correct $-1/r$ asymptotic decay, which is essential for accurately describing spatially separated charge-transfer (CT) states. \cite{Stein2009Reliable,singh2025simplifiedphysicallymotivateduniversally,Mandal2025Simplified}

The accuracy of RSH functionals depends critically on the choice of the range-separation parameter $\omega$. To balance physical accuracy and computational efficiency, we determine $\omega$ using a recently proposed effective tuning scheme, $\omega_{eff}$, which provides a physically motivated alternative to conventional ionization-potential tuning. \cite{singh2025simplifiedphysicallymotivateduniversally,new_aditi} 

Within this framework, the effective range-separation parameter is defined as
\begin{equation}
\omega_{\mathrm{eff}} =
\frac{a_1}{\langle r_s \rangle}
+ \frac{a_2 \langle r_s \rangle}{1 + a_3 \langle r_s \rangle^2},
\label{eq:omega_eff}
\end{equation}
where \(a_1 = 1.91718\), \(a_2 = -0.02817\), and \(a_3 = 0.14954\). As indicated in Eq.~\ref{eq:omega_eff}, $\omega_{eff}$ depends explicitly on the average Wigner-Seitz (WS) radius $\langle r_s \rangle$. Importantly, the parameters originate from the underlying theoretical formulation and are not empirically fitted.\cite{jana2023simple}. The quantity $\langle r_s \rangle$ is evaluated from the ground-state electron density $\rho(r)$ as defined in Eq.~\ref{eq:rs_avg}
\begin{equation}
\langle r_s \rangle =
\frac{
\displaystyle
\int
\mathrm{erf}\!\left(
\frac{\rho(\mathbf{r})}{\rho_{\mathrm{c}}}
\right)
r_s(\mathbf{r}) \,
\mathrm{d}^3\mathbf{r}
}{
\displaystyle
\int
\mathrm{erf}\!\left(
\frac{\rho(\mathbf{r})}{\rho_{\mathrm{c}}}
\right)
\mathrm{d}^3\mathbf{r}
},
\label{eq:rs_avg}
\end{equation}
where \( r_s(\mathbf{r}) \) denotes the local WS radius defined in Eq.~\ref{eq:rs_local} as
\begin{equation}
r_s(\mathbf{r}) =
\left(
\frac{3}
{4\pi \rho(\mathbf{r})}
\right)^{1/3},
\label{eq:rs_local}
\end{equation}
and \( \rho_{\mathrm{c}} \) represents a system-dependent density cutoff introduced to suppress contributions from low-density regions, as defined in Eq.~\ref{Eq:error2}
\begin{equation}
\rho_{\mathrm{c}} =
\frac{
\rho_{\mathrm{th}}
}{
\displaystyle
\int \rho(\mathbf{r}) \,
\mathrm{d}^3\mathbf{r}
},
\label{Eq:error2}
\end{equation}
where \( \rho_{\mathrm{th}} = 1.64 \times 10^{-2} \) e/bohr\(^3\). For a complete derivation and further details, readers are referred to Ref.~\citenum{singh2025simplifiedphysicallymotivateduniversally}.

Excited-state properties were evaluated using linear-response time-dependent density functional theory (TDDFT) based on the $\omega_{eff}$-tuned LC-$\omega$PBE functional. To ensure computational efficiency and numerical stability while maintaining accuracy for CT excitations, we employ the Tamm-Dancoff approximation (TDA), in which the coupling between excitation and de-excitation channels is neglected. \cite{Hirata1999} The TDA reduces the TDDFT problem to a Hermitian eigenvalue equation, enabling the use of robust and efficient numerical solvers.

Beyond its computational advantages, the TDA is known to suppress spurious low-energy instabilities that can arise in full TDDFT calculations, particularly for systems with small fundamental gaps or pronounced CT character. \cite{Magyar2007dependence,Sears2011communication,Rangel2017assessment} Moreover, TDA-TDDFT typically provides an accurate description of excitation energies dominated by single-electron transitions, which are most relevant for optical absorption and photoinduced electron-injection processes in DSSC sensitizers. \cite{Hirata1999,Casanova2020assessing}


\begin{figure*}
    \centering
    \includegraphics[width=0.9\linewidth]{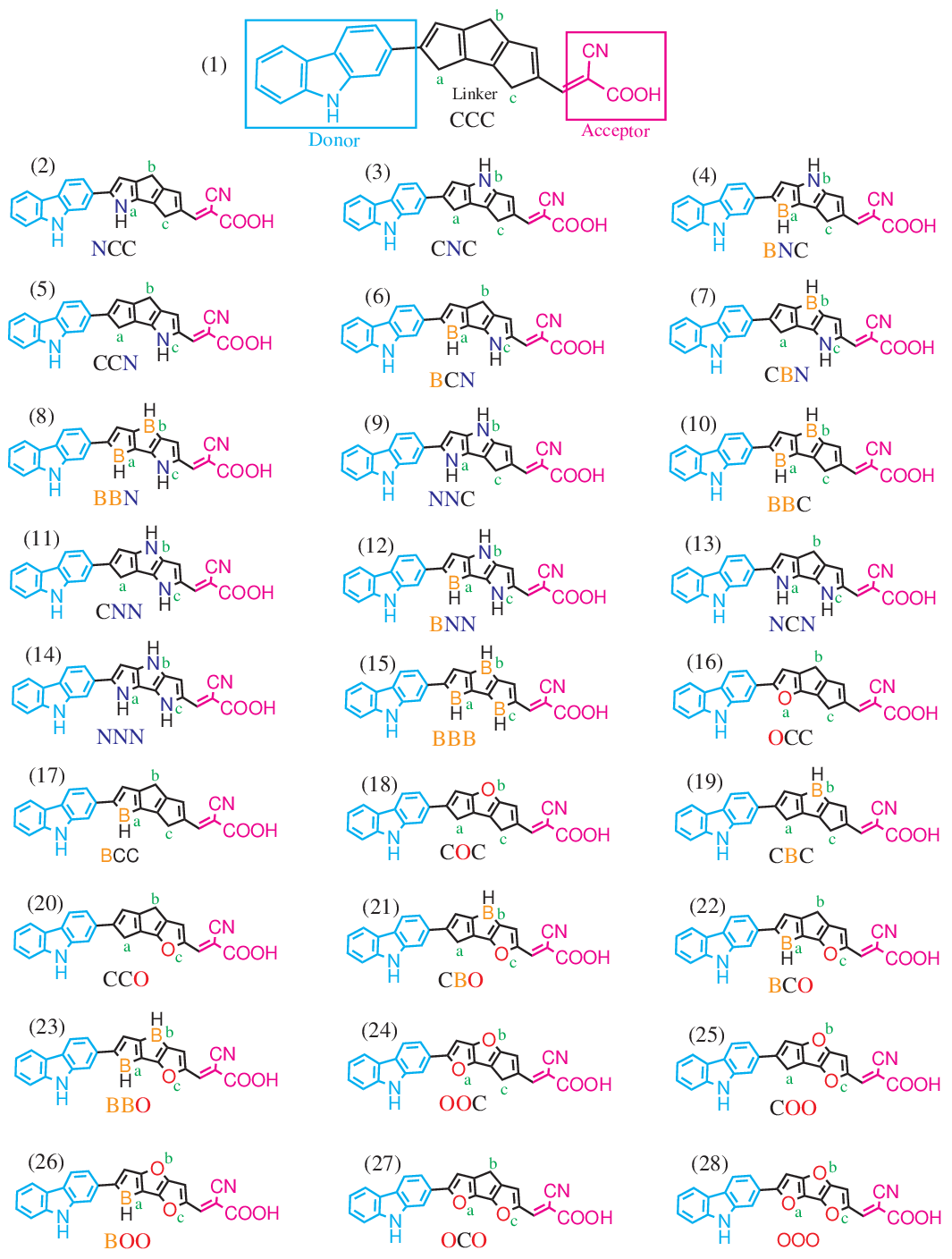}
    \caption{Schematic representation of a library comprising 27 mono-, di-, and tri-doped prototypical organic dyes featuring common donor and acceptor moieties, where a, b, and c denote the hetero-atom doping positions within the linker unit.}
    \label{fig:structure}
\end{figure*}
\begin{figure*}
    \centering
    \includegraphics[width=\linewidth]{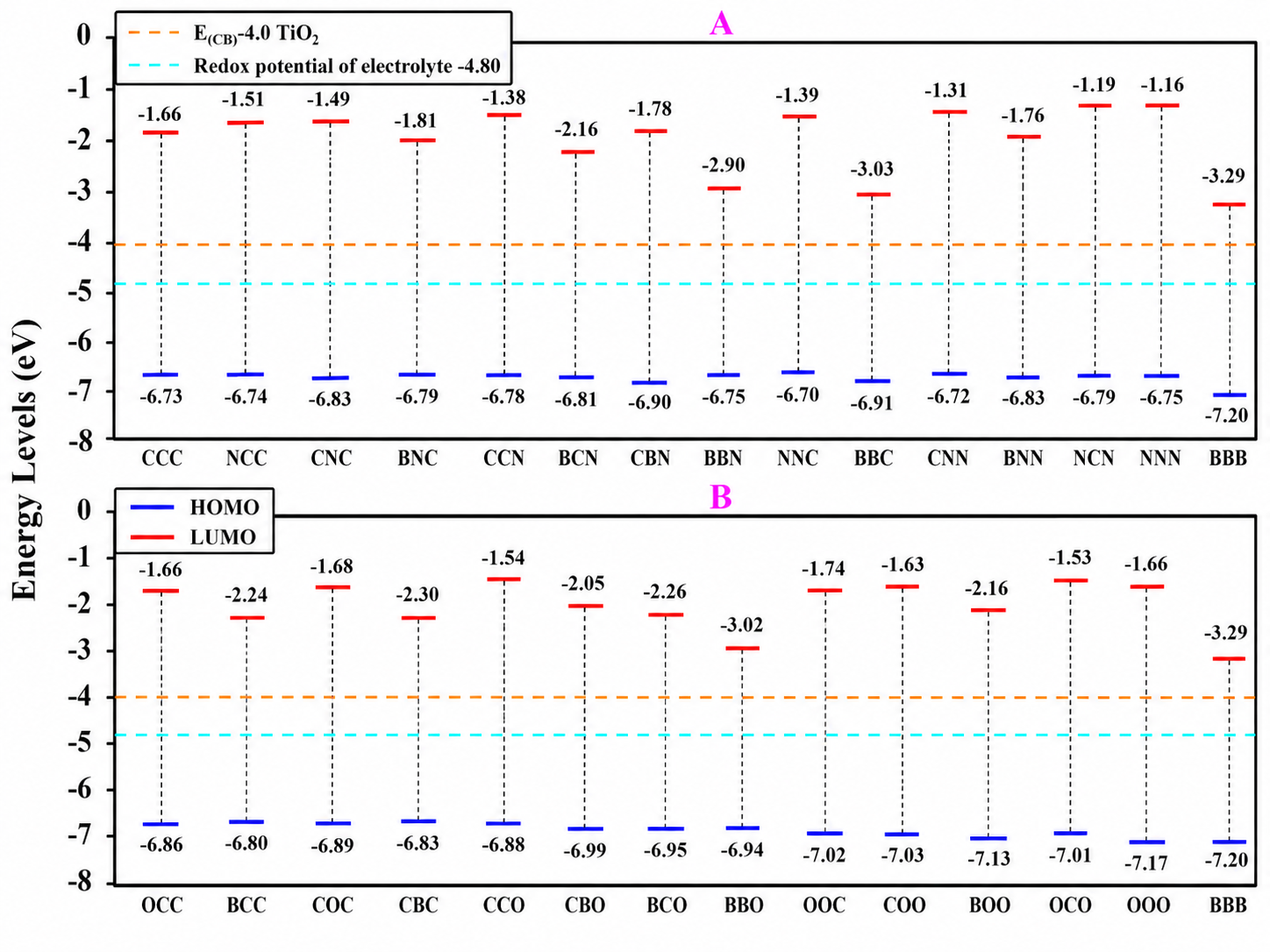}
    \caption{The HOMO and LUMO energy levels (eV) of N- and B-doped (A) and O- and B-doped (B) organic dyes, analyzing the performance with the LC-$\omega$PBE functional and def2-TZVPD basis set using the effective tuning parameter ($\omega_{eff}$). The complete data is available in the SI Table S4. For comparison to IP tuning ($\omega_{IP}$), refer to SI Figure S1 and S2.}
    \label{fig:homo-lumo-level1}
\end{figure*}


\begin{figure*}
    \centering
    \includegraphics[width=\linewidth]{doping_trend_eff.eps}
    \caption{The HOMO-LUMO gap trends for the doped system, analyzing the performance with the LC-$\omega$PBE functional and def2-TZVPD basis set using the effective tuning parameter ($\omega_{eff}$). Graphs A-G denote progressively increasing boron content in the undoped, mono-, di-, and tri-doped nitrogen, and mono-, di-, and tri-doped oxygen organic dyes, respectively. The complete data is available in the SI Table S4. For comparison to IP tuning ($\omega_{IP}$), refer to SI Figure S3.}
\label{fig:homo-lumo-gap}
\end{figure*}

\begin{figure*}
    \centering
    \includegraphics[width=0.8\linewidth]{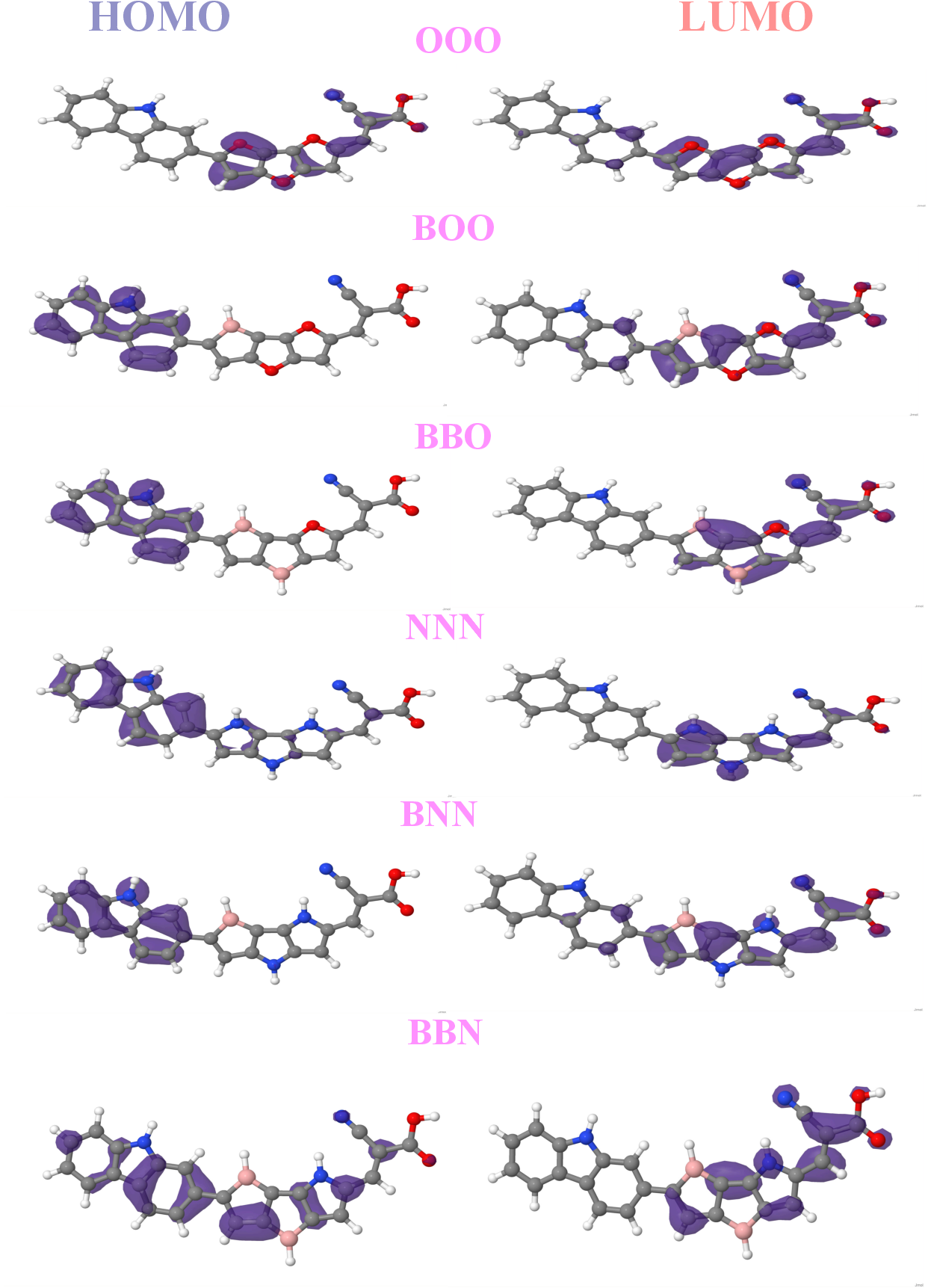}
    \caption{The spatial distributions of the frontier molecular orbitals (HOMO and LUMO) for O-, B-, and N-doped organic dyes, calculated using the LC-$\omega_{eff}$PBE functional and def2-TZVPD basis set.}
    \label{fig:homo-lumo-orb1}
\end{figure*}

\begin{figure}
    \centering
    \includegraphics[width=\linewidth]{doping_SINGLET_triplet_increasing_eff.eps}
    \caption{The singlet-singlet (A) and singlet-triplet (B) excitation energies (eV) of N-, O-, and B-doped organic dyes, analyzing the performance with LC-$\omega$PBE functional and def2-TZVPD basis set using the effective tuning parameter ($\omega_{eff}$). The complete data is available in the SI Table S5. For comparison to IP tuning ($\omega_{IP}$), refer to SI Figure S4.}
    \label{fig:ss}
\end{figure}


\subsection{B. Computational details}
The geometric structures for a series of mono-, di-, and tri-doped organic dyes were optimized in a vacuum using KS-DFT with the BP86~\cite{bp86-becke,bp86-perdew} functional and the cc-pVDZ basis set,~\cite{Dunning} using \textsc{orca} software package~\cite{ORCA5}. All optimized structures were confirmed as true energy minima by the absence of imaginary frequencies in subsequent vibrational frequency analysis. The corresponding xyz coordinates for all structures are available in Ref.~\cite{dopedsystems} \\
Electronic structure calculations were conducted within the framework of KS-DFT using the long-range-corrected LC-$\omega$PBE functional in combination with the Def2-TZVPD basis set for all molecular systems. This approach enabled the accurate determination of frontier molecular orbital energies, including the HOMO and LUMO orbitals, as well as the corresponding HOMO-LUMO energy gaps. Excited-state properties were subsequently computed using TDA to obtain vertical excitation energies. All electronic structure calculations were carried out using the Q-Chem\cite{qchem} software package. The resulting frontier molecular orbitals were visualized and analyzed using the Jmol software package\cite{jmol} to examine orbital distributions, and spatial localization features relevant to CT processes.\\
The $\omega_{eff}$ parameter have been computed using PySCF~\cite{pyscf} with a publicly available tuning protocol\cite{repo}.This protocol employs an averaging approach that exhibits minimal sensitivity to variations in functional or basis set selection.\cite{singh2025simplifiedphysicallymotivateduniversally} Moreover, for comparison we have also determined the optimal $\omega$ parameter using IP-tuning procedure as
\begin{equation}
\omega_{IP} = \arg\min_{\omega} \left| \mathrm{IP}(\omega) + \varepsilon_{\mathrm{HOMO}}(\omega) \right| \; .
\end{equation}

Although this tuning procedure may fail in some cases for CT systems\cite{failCT2} the comparison with effective tuning is important to validate the whole computational protocol. 
In this case the range-separation parameter $\omega$ is optimized to satisfy the ionization potential (Koopmans' theorem) condition\cite{Kronik2012excitation,Refaely2012Quasiparticle}. This step is done in order to systematically compare the robustest of effective tuning. The IP-tuned parameter is determined via Q-Chem software\cite{qchem}, with aug-cc-pVDZ\cite{aug-cc-pvdz} basis and LC-$\omega$PBE functional for optimization. 
 This basis set has been extensively utilized in several previous investigations\cite{vikram2018impact, Kronik2012excitation, herbert2008simultaneous} and has been shown to introduce only minor changes relative to larger basis sets, while still providing reliable results at a lower computational cost.

\subsection{C. Doped systems}
We constructed a comprehensive library of doped organic dyes by selectively replacing carbon atoms with heteroatoms ({\color{blue}N}, {\color{red}O}, {\color{orange}B}) at three critical linker sites (denoted as positions \textit{a}, \textit{b}, and \textit{c}). The library was systematically designed to include mono-, di-, and tri-doped configurations. For mono-doping, we designed three isomers for both nitrogen ({\color{blue}N}CC, C{\color{blue}N}C, CC{\color{blue}N}) and oxygen ({\color{red}O}CC, C{\color{red}O}C, CC{\color{red}O}) and two isomers for boron ({\color{orange}B}CC and C{\color{orange}B}C), yielding a total of eight mono-doped systems. Di-doping yielded both homogeneous ({\color{blue}N}{\color{blue}N}C, C{\color{blue}N}{\color{blue}N}, {\color{blue}N}C{\color{blue}N}, {\color{red}O}{\color{red}O}C, C{\color{red}O}{\color{red}O}, {\color{red}O}C{\color{red}O}, and {\color{orange}B}{\color{orange}B}C) and mixed-heteroatom ({\color{orange}B}{\color{blue}N}C, {\color{orange}B}C{\color{blue}N}, C{\color{orange}B}{\color{blue}N}, {\color{orange}B}C{\color{red}O}, and C{\color{orange}B}{\color{red}O}) configurations, giving a total of twelve di-doped systems. The tri-doped set includes both pure ({\color{blue}N}{\color{blue}N}{\color{blue}N}, {\color{red}O}{\color{red}O}{\color{red}O}, {\color{orange}B}{\color{orange}B}{\color{orange}B}) and mixed configurations ({\color{orange}B}{\color{orange}B}N, B{\color{blue}N}{\color{blue}N}, {\color{orange}B}{\color{orange}B}O, B{\color{red}O}{\color{red}O}), providing a total of seven tri-doped systems. Finally, this results in an overall total of 27 doped configurations which are depicted in Figure~\ref{fig:structure}.\\
A systematic naming protocol was established where the sequence of letters in the dye's name corresponds to the atom type at each position. For instance, a dye with nitrogen ({\color{blue}N}) at position \textit{a} and carbon (C) at positions \textit{b} and \textit{c} is denoted {\color{blue}N}CC, while a dye doped with three nitrogen atoms at positions \textit{a, b,} and \textit{c} is named {\color{blue}N}{\color{blue}N}{\color{blue}N}. This scheme can be applied for mapping other systems, as illustrated in Figure~\ref{fig:structure}, which provides a clear and consistent label for all variants.

\section{III. Results and Discussions:}
This work investigates a series of newly designed BN-doped (boron and nitrogen), BO-doped (boron and oxygen), and boron-doped $\pi$-conjugated organic dyes with a common donor and acceptor. Nitrogen- and oxygen-doped dyes are included from the literature~\cite{ram-jpca,jacsexperimental}, as depicted in Figure~\ref{fig:structure}.
These dyes are designed by doping heteroatoms ({\color{blue}N}, {\color{red}O}, and {\color{orange}B}) at critical positions within the $\pi$-conjugated linker or bridge part of the undoped reference dye (CCC). 
The 
linker connects a carbazole donor to a cyanoacrylic acid acceptor. The whole molecular backbone exhibits an alternating pattern of single and double bonds (known as conjugation). However, one specific carbon atom in each ring of the linker is excluded from this conjugated pathway. Our objective is to dope these critical positions of carbon, labeled \textit{a, b,} and \textit{c}, with heteroatoms. This transformation represents an effective strategy to extend the conjugation and enhance delocalization of the electron across the entire system, which helps in facilitating efficient CT in DSSCs.~\cite{ct5,ct1,efficient-charge-transfer}
\begin{table}[h!]
\centering
\small
\rowcolors{2}{cyan!20}{white}
\begin{tabular}{c|c|c|c|c}

Molecule  & CCSD(T)&LC-$\omega_{eff}$PBE & LC-$\omega_{IP}$PBE   & {Exp.} \\ 
\hline

{{\color{orange}B}{\color{blue}N}-1,2 Naph} &8.41&8.46&8.36&{8.45} \\

{{\color{orange}B}{\color{blue}N}-1,9 Naph}  & 7.79 &7.87 &7.73&{7.78} \\

{\color{orange}B}{\color{blue}N}-9,1-Naph  & 7.47&7.48&7.38&{7.44} \\ 

{\color{orange}B}{\color{blue}N}-9,10-Naph &8.19&8.32&8.29&{8.42} \\
\hline 

{MAE (Exp.)} &0.08&0.06 &0.08\\

\end{tabular}
\caption{Vertical ionization potentials in eV calculated using the LC-$\omega$PBE functional with the cc-pVTZ basis set in Q-Chem\cite{qchem}. The CCSD(T) calculations were carried out using the MOLPRO software package\cite{molpro}. The geometries of mono-BN-doped naphthalene are from Ref.\citenum{ram2025frozen}, and the UV-visible spectroscopy experimental results are extracted from Ref.\citenum{jacsexperimental,ram2025frozen}.  Mean absolute error (MAE) with respect to experimental values.
}
\label{tab:experiment}
\end{table}

The primary objective of this work is to deliver reliable predictions of key properties for newly designed heteroatom-doped prototypical dyes, specifically frontier orbital energy levels and singlet-singlet (SS) and singlet-triplet (ST) excitation energies, using an effective tuned DFT/TDDFT protocol across a series of novel organic dyes.

\subsection{A. Initial validation}

In order to perform initial validation of this method for doped molecular systems, we applied the $\omega_{eff}$ RHS computational protocol to 
four heteroatom-doped organic molecules found in the literature~\cite{jacsexperimental,ram2025frozen}, namely BN-1,2-naphthalene, BN-1,9-naphthalene, BN-9,1-naphthalene, and BN-9,10-naphthalene isomers for which experimental vertical IPs (VIPs) data are available. For comparison, we computed the VIPs using the highly accurate CCSD(T) (coupled cluster with single, double, and perturbative triple excitations) method in the cc-pVTZ basis set. These are reported in \Tab{tab:experiment} together with LC-$\omega$PBE result obtained for $\omega_{eff}$ and $\omega_{IP}$ tuning procedure.

The results indicates that the optimally tuned LC-$\omega$PBE approaches deliver consistently high accuracy for VIPs, reaching near-chemical accuracy ($\approx$ 0.1 eV) at a fraction of the cost of CCSD(T). In particular, the LC-$\omega_{eff}$PBE variant provides the best overall agreement with experiment, yielding the lowest mean absolute error among the tested methods, and showing no large systematic failures across the dataset. The LC-$\omega$PBE accuracy has been also confirmed in similar context in Ref. \citenum{Rodrigo}.

Although the CCSD(T) performs very accurately in this small set, it does not clearly outperform the best-tuned DFT protocol in terms of average deviation from experiment. This underscores an important practical point, namely, for ionization energies in these extended $\pi$-systems, well-tuned RSHs can match (or even slightly surpass) CCSD(T) on average, while being far more computationally affordable. 

Comparing the two tuning strategies, LC--$\omega_{\mathrm{eff}}$PBE and LC--$\omega_{\mathrm{IP}}$PBE exhibit differences in the overall error statistics that remain within the uncertainty of the reference data. Given this comparable performance, the computationally more economical LC--$\omega_{\mathrm{eff}}$PBE approach provides an attractive and efficient alternative. The latter variant shows a slightly larger average deviation, consistent with a mild systematic bias relative to the experiment. 

We further note that the parameter $\omega_{eff}$ has been extensively validated, demonstrating its ability to reproduce experimental VIPs, to deliver HOMO-LUMO gaps comparable in quality to Green's function methods, and to accurately predict SS excitation energies in organic photovoltaic molecules (see Ref.~\cite{singh2025simplifiedphysicallymotivateduniversally,new_aditi}).
Taken together, the results support the claim that effective tuning is a robust and efficient protocol and seems to be an ideal tool for exploring novel organic electronic systems.

\subsection{B. HOMO-LUMO Gap}
The objective of this work is to precisely control the energy and character of the frontier molecular orbitals (HOMO/LUMO) through targeted doping of the bridge of the core structure of undoped dye (CCC) at specific positions \textit{a, b,} and \textit{c} with boron, oxygen, and nitrogen atoms, as shown in Figure~\ref{fig:structure}. The effective-tuning
methodology is used to compute the HOMO and LUMO energy levels of all newly designed prototypical organic dyes, which play a crucial role in predicting optoelectronic characteristics and CT in DSSCs.~\cite{homo-lumo-play-crucial-role,ct4} Comparison with IP-tuned results is provided in the Supporting Information (SI) \cite{supplementary}(see Figure S1 and S2). Importantly, a strong agreement is observed, as the trends are comparable and the values show minimal deviation between the two tuning procedures. These energy levels are well-aligned in a way that is highly compatible with the key components of the cells, namely the conduction band energy (orange dashed line) and the electrolyte redox potential (cyan dashed line), as depicted in Figure~\ref{fig:homo-lumo-level1}A and Figure~\ref{fig:homo-lumo-level1}B. For proper device function, the HOMO must be energetically below the redox potential of the electrolyte to enable dye regeneration, while the LUMO must be above the conduction band of the semiconductor to facilitate electron injection. This energetic alignment demonstrates that all dyes possess the characteristics required for a sensitizer and can be considered promising candidates for efficient CT in DSSCs. This also shows that the $\omega_{eff}$ RS computational protocol is able to provide relatively fast tools for initial validation of new DSSC materials.

The sequential substitution of carbon with boron (CCC$\rightarrow${\color{orange}B}CC$\rightarrow${\color{orange}B}{\color{orange}B}C) at positions \textit{a} and \textit{b} of bridge in the core molecular structure of the undoped organic dye (CCC) substantially decreases the HOMO-LUMO gap, reflecting the electron-accepting ability of boron from the adjacent electron donating donor moiety and enhanced $\pi$-conjugation,~\cite{decrease-homo-lumo-gap1,decrease-homo-lumo-gap2} which is crucial for CT.~\cite{ct2,ct3} However, this gap increases slightly when the boron is doped at all three positions, namely, \textit{a, b,} and \textit{c} to form ({\color{orange}B}{\color{orange}B}{\color{orange}B}). The observed increase in the gap arises due to incorporation of an electron-deficient boron atom at the terminal position \textit{c}, which is located too far from the electron-donating donor, and therefore disrupts the $\pi$-electron delocalization that helps to lower the HOMO-LUMO gap, as shown in Figure~\ref{fig:homo-lumo-gap}A.\\
In mono-doped ({\color{blue}N}CC, C{\color{blue}N}C, and CC{\color{blue}N}) and di-doped ({\color{blue}N}{\color{blue}N}C, C{\color{blue}N}{\color{blue}N}, and  {\color{blue}N}C{\color{blue}N}) nitrogen-based organic dyes, the HOMO-LUMO gap increases as the position of nitrogen atom moves from ({\color{blue}N}CC$\rightarrow$C{\color{blue}N}C$\rightarrow$CC{\color{blue}N}) and ({\color{blue}N}{\color{blue}N}C$\rightarrow$ C{\color{blue}N}{\color{blue}N}$\rightarrow$ {\color{blue}N}C{\color{blue}N}), respectively, as depicted in Figure~\ref{fig:homo-lumo-level1}A. This gap is tunable and can be substantially decreased through the progressive incorporation of boron at positions \textit{a} and \textit{b} (CC{\color{blue}N}$\rightarrow${\color{orange}B}C{\color{blue}N}$\rightarrow${\color{orange}B}{\color{orange}B}{\color{blue}N}, C{\color{blue}N}{\color{blue}N}$\rightarrow${\color{orange}B}{\color{blue}N}{\color{blue}N}$\rightarrow${\color{orange}B}{\color{orange}B}{\color{blue}N}) as shown in Figure~\ref{fig:homo-lumo-gap}B and ~\ref{fig:homo-lumo-gap}C, respectively.\\
In mono-doped ({\color{red}O}CC, C{\color{red}O}C, and CC{\color{red}O}) and di-doped ({\color{red}O}{\color{red}O}C, C{\color{red}O}{\color{red}O}, and  {\color{red}O}C{\color{red}O}) oxygen-based dyes, the HOMO energies are nearly identical. However, the LUMO energies increases slightly when the position of the oxygen atom changes ({\color{red}O}CC$\rightarrow$C{\color{red}O}C$\rightarrow$CC{\color{red}O}) and ({\color{red}O}{\color{red}O}C$\rightarrow$ C{\color{red}O}{\color{red}O}$\rightarrow$ {\color{red}O}C{\color{red}O}), as shown in Figure~\ref{fig:homo-lumo-level1}B. As a result, the observed HOMO-LUMO gap increases across the series. In contrast, this gap decreases significantly with incorporation of boron doping (CC{\color{red}O}$\rightarrow${\color{orange}B}C{\color{red}O}$\rightarrow${\color{orange}B}{\color{orange}B}{\color{red}O}, C{\color{red}O}{\color{red}O}$\rightarrow${\color{orange}B}{\color{red}O}{\color{red}O}$\rightarrow${\color{orange}B}{\color{orange}B}{\color{red}O}) at positions \textit{a} and \textit{b}, as illustrated in Figure~\ref{fig:homo-lumo-gap}E and ~\ref{fig:homo-lumo-gap}F, respectively. The collective information from both approaches demonstrates a consistent trend: mono- and di-doping with heteroatoms ({\color{blue}N} and {\color{red}O}) increases the HOMO-LUMO gap across the series, whereas the sequential introduction of boron atoms significantly decreases it, as shown in Figure S1, S2 and S3 in the SI.\\
In the tri-doped systems ({\color{blue}N}{\color{blue}N}{\color{blue}N}, {\color{red}O}{\color{red}O}{\color{red}O}, and {\color{orange}B}{\color{orange}B}{\color{orange}B}), the HOMO and LUMO energy levels in the oxygen- and nitrogen-based systems are at different positions. However, their HOMO-LUMO gaps are almost the same. The HOMO-LUMO gap of these two systems is higher than that of the purely  boron-based system ({\color{orange}B}{\color{orange}B}{\color{orange}B}). This gap can be systematically reduced by increasing the boron content through atomic substitution at the bridge positions \textit{a} and \textit{b}, as illustrated in Figure~\ref{fig:homo-lumo-gap}D and ~\ref{fig:homo-lumo-gap}G for the sequences ({\color{blue}N}{\color{blue}N}{\color{blue}N}$\rightarrow${\color{orange}B}{\color{blue}N}{\color{blue}N}$\rightarrow${\color{orange}B}{\color{orange}B}{\color{blue}N}) and ({\color{red}O}{\color{red}O}{\color{red}O}$\rightarrow${\color{orange}B}{\color{red}O}{\color{red}O}$\rightarrow${\color{orange}B}{\color{orange}B}{\color{red}O}), respectively. Figure~\ref{fig:homo-lumo-orb1} displays the spatial distribution of these frontier orbitals; thus, it can be observed that the reduction in the electronic gap is correlated with a progressive shift of the HOMO orbital from the bridge and acceptor parts of the organic dye to the donor part. However, the LUMO orbital remains localized over the bridge and acceptor part and is largely intact. This is observed upon sequential incorporation of boron at positions \textit{a} and \textit{b} in both the {\color{blue}N}{\color{blue}N}{\color{blue}N} and {\color{red}O}{\color{red}O}{\color{red}O} parent systems, forming the {\color{orange}B}{\color{blue}N}{\color{blue}N}/{\color{orange}B}{\color{orange}B}{\color{blue}N} and {\color{orange}B}{\color{red}O}{\color{red}O}/{\color{orange}B}{\color{orange}B}{\color{red}O} systems, respectively. The smallest gap among all variants is observed in the {\color{orange}B}{\color{orange}B}{\color{blue}N} dye, which is doped with two boron ({\color{orange}B})  atoms at positions \textit{a} and \textit{b} and one nitrogen ({\color{blue}N}) atom at position \textit{c} on the bridge.\\

\subsection{C. Singlet-Singlet and Singlet-Triplet Excitation Energies}
The general trends of SS and ST excitation energies for the series of doped dyes are presented systematically in increasing order in Figure~\ref{fig:ss}A and~\ref{fig:ss}B, respectively. These excitation energies are obtained using both effective and IP-tuning methodologies (see Figure S4 in SI file). Both tuning procedures predict the same overall behavior, differing only slightly in a few isolated cases, which supports our analysis. The undoped organic dye (CCC) is used as the reference for interpreting the results. Overall, the data reveal two distinct, opposing trends controlled by the electronic character of the heteroatom dopants-specifically, whether the dopant acts as an electron donor or acceptor. Incorporation of electron-rich, lone-pair-bearing heteroatoms such as nitrogen and oxygen markedly increases the excitation energies relative to CCC. This increase is larger for nitrogen-doped systems than for their oxygen-doped counterparts. The corresponding blue shift is consistent across mono-, di-, and tri-doped N- and O-containing dyes. Moreover, the excitation energy increases systematically with dopant count ({\color{blue}N} or {\color{red}O}): tri-doped dyes exhibit higher energies than di-doped dyes, which in turn exceed those of mono-doped dyes.

In contrast, the incorporation of the electron-deficient boron ({\color{orange}B}) atom is a highly effective strategy for tuning the optical properties, as it systematically and significantly reduces this excitation energy, leading to a pronounced red-shift. This effect follows a clear concentration dependence across all doping levels, as demonstrated by the progressive decrease in SS and ST excitation energies along sequences in mono-doped (CC{\color{blue}N}$\rightarrow${\color{orange}B}C{\color{blue}N}$\rightarrow${\color{orange}B}{\color{orange}B}{\color{blue}N}, CC{\color{red}O}$\rightarrow${\color{orange}B}C{\color{red}O}$\rightarrow${\color{orange}B}{\color{orange}B}{\color{red}O}), di-doped (C{\color{blue}N}{\color{blue}N}$\rightarrow${\color{orange}B}{\color{blue}N}{\color{blue}N}$\rightarrow${\color{orange}B}{\color{orange}B}{\color{blue}N}, C{\color{red}O}{\color{red}O}$\rightarrow${\color{orange}B}{\color{red}O}{\color{red}O}$\rightarrow${\color{orange}B}{\color{orange}B}{\color{red}O}), and tri-doped ({\color{blue}N}{\color{blue}N}{\color{blue}N}$\rightarrow${\color{orange}B}{\color{blue}N}{\color{blue}N}$\rightarrow${\color{orange}B}{\color{orange}B}{\color{blue}N}, {\color{red}O}{\color{red}O}{\color{red}O}$\rightarrow${\color{orange}B}{\color{red}O}{\color{red}O}$\rightarrow${\color{orange}B}{\color{orange}B}{\color{red}O}) systems,  as illustrated in Figure~\ref{fig:ss}. For ST excitation we observe negative excitation energies for {\color{orange}B}{\color{orange}B}{\color{blue}N}, {\color{orange}B}{\color{orange}B}{\color{red}O} and C{\color{orange}B}C systems. 
This reveals their diradical character and indicates a strong instability toward a triplet solution, suggesting significant polyradical singlet nature. This observation is consistent with earlier studies reported for TDDFT\cite{ Sears2011communication, Felixmeaning2025}, as well as for WFT and multireference methods\cite{sanz2021negative, lucie2022origin, Ghosh2022origin}.

Among all newly designed dyes, the {\color{orange}B}{\color{orange}B}{\color{blue}N} dye exhibits the lowest SS and ST excitation energies. 
\section{IV. Conclusion}

CT dye discovery for DSSCs requires exploring a large chemical space while preserving predictive accuracy for frontier-orbital alignment and low-lying excited states. In practice, this creates a central bottleneck: high-level WFT methods are often very expensive for systematic screening, while conventional tuning of RSH functionals (e.g., IP tuning) adds substantial overhead that limits throughput and dataset scale. Addressing this challenge, we analyzed and deployed a novel single-shot effective-tuning protocol, $\omega_{eff}$, as a simple and cost-effective alternative to IP tuning for determining the optimal RS parameter in LR corrected hybrid functionals, while maintaining accuracy close to WFT benchmarks. 

A key outcome of this work is that the availability of a reliable, low-cost tuning strategy enables \emph{effective screening} and, therefore, rational \emph{design} of new CT candidates. Specifically, the validated protocol $\omega_{eff}$ allowed us to systematically construct and characterize an organized benchmark/library set of 27 mono-, di-, and tri-doped organic dyes based on the D-$\pi$-A model consisting of a common carbazole donor and cyanoacrylic acid acceptor, with targeted substitutions at three critical bridge sites (\textit{a}, \textit{b}, and \textit{c}) using N, O, and B dopants. This dataset provides a consistent platform for extracting clear structure property relationships and guiding future candidate selection in a way that would be significantly more difficult with costlier tuning procedures or WFT-based screening.

Across this doped dye library, we show that electronic and optical properties can be strongly tuned through strategic heteroatom doping at the bridge positions. Doping with electron-rich nitrogen or oxygen generally increases the HOMO--LUMO gap as well as the SS and ST excitation energies, with a markedly stronger effect for nitrogen than for oxygen. Moreover, for N- and O-containing series, the excitation energies increase progressively with the number of dopants, following the order: mono-doped $<$ di-doped $<$ tri-doped dyes, consistent with systematic blue shifts. In contrast, the sequential incorporation of electron-deficient boron at these sites substantially reduces the HOMO-LUMO gap and lowers both SS and ST excitation energies, producing pronounced red shifts. Fundamentally, these opposing trends stem from the donor versus acceptor character of the dopants and their impact on $\pi$-electron delocalization and frontier orbital localization within the donor-bridge-acceptor framework. Collectively, these results establish boron incorporation as a powerful strategy for engineering sensitizers with reduced gaps and lower excitation energies, which are key ingredients for efficient light harvesting and charge separation in DSSCs.

More broadly, this work demonstrates the expanded utility of LR-corrected hybrid functionals for materials design and highlights $\omega_{eff}$ as a computationally efficient and accessible pathway to accurate predictions relevant to CT. By lowering the cost barrier to systematic exploration, the effective-tuning protocol enables the creation of curated benchmark libraries and supports the rapid screening and rational design of novel donor-acceptor architectures. Future work will extend this framework to more device-realistic modeling, including solvent and interfacial effects, explicit dye-\ce{TiO2} binding motifs, and additional CT descriptors, to further connect molecular-level screening to experimentally measurable DSSC performance metrics.

\section*{Acknowledgements}
S.\'S. acknowledges the financial support from the National Science Centre, Poland (grant no. 2021/42/E/ST4/00096). 
R. D. P. gratefully acknowledges financial support from the Polish National Agency for Academic Exchange (NAWA).
R.~D.~P. and P.~T.~acknowledge financial support from the PRELUDIUM BIS research grant from the National Science Centre, Poland (Grant No. 2023/50/O/ST4/00353). The computational work for this study was performed during a three-month research visit at the National Institute of Science Education and Research (NISER) in Bhubaneswar, India.
The authors gratefully acknowledge the use of the HPC cluster Kalinga, located in the School of Physical Sciences at NISER.

\section*{Data Availability}
The data that support the findings are published within this study.

\section*{Supporting Information}

\subsection*{1. Supporting Information Data}

The Supporting Information file includes the following data for BN-doped and other heteroatom-doped carbazole-based systems:

\begin{itemize}
  \item Values of the range-separation parameters $\omega_{\mathrm{eff}}$ and $\omega_{\mathrm{IP}}$.
  \item Frontier molecular orbital energies (HOMO and LUMO) and the corresponding fundamental HOMO-LUMO gaps.
  \item Vertical excitation energies corresponding to singlet-singlet and singlet-triplet excited states.
\end{itemize}

\subsection*{2. Molecular Geometry Data}

The optimized molecular geometries of all heteroatom-doped carbazole-based systems are available via the following public GitHub repository:
\url{https://github.com/aditisingh4812/Doped-systems}

\section*{Conflicts of interest}
There are no conflicts to declare.

\bibliography{reference}
\end{document}